\documentclass[aps,prd,preprint,tightenlines,
   showpacs,nofootinbib]{revtex4}
\usepackage{graphicx}
\usepackage{dcolumn}
\usepackage{bm}
\newcommand{\beq}{\begin{equation}}
\newcommand{\eeq}{\end{equation}}
\newcommand{\beqa}{\begin{eqnarray}}
\newcommand{\eeqa}{\end{eqnarray}}

\def\lta{~\mbox{\raisebox{-.6ex}{$\stackrel{<}{\sim}$}}~}
\def\gta{~\mbox{\raisebox{-.6ex}{$\stackrel{>}{\sim}$}}~}

\def\beq{\begin{equation}}
\def\eeq{\end{equation}}
\def\bea{\begin{eqnarray}}
\def\eea{\end{eqnarray}}
\newcommand{\be}{\begin{equation}}

\newcommand{\ee}{\end{equation}}
\def\beqa{\begin{eqnarray}}
\def\eeqa{\end{eqnarray}}
\def\beq{\begin{equation}}
\def\eeq{\end{equation}}

\let\gam=w

\renewcommand{\epsilon}{\varepsilon}

\newcommand{\gev}{\text{GeV}}

\def\su{{\rm SUSY}}

\begin{document}

\preprint{FTPI-MINN-09/19}
\preprint{UMN-TH-2747/09}

\vspace{8cm}

\title{Quantum Corrections to the Cosmological Evolution of Conformally Coupled Fields}

\author{Jose A. R. Cembranos$^{1,2}$, Keith A. Olive$^{1,2}$, Marco Peloso$^2$
and Jean-Philippe Uzan$^3$}

\affiliation{
${^1}$ William I. Fine Theoretical Physics Institute,
University of Minnesota, Minneapolis, 55455, USA \\
${^2}$ School of Physics and Astronomy,
University of Minnesota, Minneapolis, 55455, USA \\
$^3$ Institut d'Astrophysique de Paris,
              UMR-7095 du CNRS, Universit\'e Pierre et Marie
              Curie,
              98 bis bd Arago, 75014 Paris (France)
}
\vspace*{2cm}

\begin{abstract}
Because the source term for the equations of motion of a conformally coupled scalar field,
such as the dilaton, is given by the trace of the matter energy momentum tensor, it is commonly assumed to vanish during 
the radiation dominated epoch in the early universe. 
As a consequence, such fields are generally
frozen in the early universe.  Here we compute the finite temperature radiative correction to
the source term and discuss its consequences on the evolution of such fields in the early universe.
We discuss in particular, the case of scalar tensor theories of gravity which have 
general relativity as an attractor solution. We show that, in some cases, the universe can experience an early phase of contraction, followed by a non-singular bounce, and standard expansion. This can have interesting consequences for the abundance of thermal relics; for instance, it can provide a solution to the gravitino problem. We conclude by discussing the possible consequences of the quantum corrections to the evolution of the dilaton.

\end{abstract}


\maketitle


\section{Introduction}
Although fundamental scalar fields have yet to be discovered, their role in fundamental
physics is unquestionable.  One particular class of scalar fields
are those which are conformally coupled to matter. Well known examples
of such fields are the Jordan-Fierz-Brans-Dicke (JFBD) scalar which is introduced as an extension
to general relativity (GR) \cite{stgen} or the dilaton which in the context of string theory arises as part of
the gravitational multiplet \cite{polchy}. Of course string theories possess many other scalars
or moduli which are conformally coupled to matter as well.  

Through a series of field redefinitions, theories with conformally coupled scalar fields (CCSFs)
can always be reexpressed in terms of Einstein gravity with well determined couplings
to the matter sector of the theory.  These new scalar interactions may be perceived as new
forces which can lead to violations of the equivalence principle if all couplings are not universal
 \cite{ctes}. Placed in a cosmological context, the evolution of these fields
can lead to the variation of fundamental constants, including gauge couplings
and Yukawa couplings in the general case. Theories with universal couplings to matter \cite{dnord,dpol} 
as well as so-called chameleon theories \cite{brax}, both of which are based on CCSFs, are constructed 
to minimize these violations (in particular they ensure the universality
of free fall and the constancy of all non-gravitational constants). 
Indeed, precision tests  of gravity often lead to important constraints on the parameters of the theory with 
these scalar fields. Cosmologically, the presence of a new dynamical degrees of freedom
can also have severe consequences on big bang nucleosynthesis (BBN) 
\cite{speedup1,bbnJFBD,bbnst1,bbn_quad,dp,couv}.
Thus there are a whole host of observations, including cosmic microwave
background anisotropies~\cite{ru02} and weak-lensing~\cite{carlo},  which 
constrain theories with CCSFs. 

The cosmological evolution of CCSFs are determined by the standard equation of motion 
for a scalar field
with an extra source term proportional to the trace of the energy momentum tensor. 
That is, in the Einstein frame, we can write (see Section \ref{cosmology-eqs} below
for details and definitions)
\beq
\ddot \phi_* + 3 H \dot \phi_* = - {dV(\phi_*) \over d\phi_*} - 4 \pi G_* \alpha(\phi_*)  T^\mu_\mu
\label{eom}
 \eeq
where star denotes quantities in the Einstein frame.
\beq
\label{alpha}
\alpha(\phi_*) \equiv {d\ln A \over d \phi_*}
\label{defalpha}
\eeq
characterizes the strength of the scalar interaction and $A(\phi_*)$ is the coupling function,
that appears
in the conformal transformation relating the string or Jordan frame to the Einstein frame
\beq 
g_{\mu\nu} = A^2(\phi_*) {g_*}_{\mu\nu}\,.
\eeq

One can recover Einstein relativity in a simple way by introducing a potential $V$ that
generates a sufficiently high mass for $\phi_*$ and forces $\phi_*$ to a constant value
(note, however, that this value will depend on the local energy density as in
the chameleon mechanism \cite{brax}). 
Even in the absence of a potential, 
this theory will have GR as an attractor if there is a well 
defined minimum to $\ln A$ \cite{dnord}. In a JFBD theory, $A\propto e^{-\lambda\phi_*}$
and there is no attraction to GR, as in the case of the dilaton in heterotic string theory.
In these models, $\alpha$ is constant and the deviation from GR is fixed (in terms of the PPN parameters). In this case, the theory is only compatible with GR if $\lambda$ is small enough. It is also clear from  
Eq.(\ref{eom}) that during the radiation dominated epoch in the early universe, the field will remain constant until $ T^\mu_\mu \ne 0$, which normally would correspond to the onset of matter domination \cite{condil}. 
This effect has been utilized, for example, as a potential solution
to the problem of a run away dilaton \cite{Brustein:2004jp}.

Assuming a perfect fluid form for the energy momentum tensor, its trace  
is simply, $\rho - 3p$, and vanishes in a radiation dominated epoch when 
the equation of state is characterized by $p = \rho/3$. While this is certainly a good
approximation deep in the radiation era, it breaks down whenever, the temperature
approaches a mass threshold, and some particle species become non-relativistic \cite{dnord,dp,brax,couv}.  There, the trace becomes proportional to
\begin{eqnarray}
\label{threshold}
 \Sigma(T) &=& \frac{g}{2\pi^2}z^2\int_{z}^\infty
 \frac{\sqrt{x^2-z^2}}{\hbox{e}^x \pm 1}dx,
\end{eqnarray}
where $g$ is the number of degrees of freedom for the particle at threshold, 
the sign refers to different statistics ($+$ for fermions, $-$ for bosons),
$T$ is the temperature of the radiation in the Jordan frame, and $z = m/T$. 
The source terms from particle thresholds and
in particular that of the electron, was shown to greatly relax the bounds from BBN on
the initial conditions of theories of gravity with CCSFs \cite{couv}.

A non-vanishing source term also arises when one includes the contribution from the trace
anomaly \cite{brax,dkkms}.  Although this is generally small - we show that $w - 1/3 = {\mathcal O} \left( 10^{-4}
- 10^{-3} \right)$ - there can be situations in which this contribution has a significant impact on the cosmology of these models. In fact, not only can it affect
the evolution of the scalar field, but we show that, in some cases, it can lead to a brief
early phase of contraction followed by a non-singular bounce in the cosmological scale factor.
In these cases, one can also derive a maximum temperature of the universe which
may provide a solution to the gravitino problem or other unwanted relics. 

The paper will be organized as follows.  In the 
next Section, we will first calculate the contribution to the equation of motion
from the trace anomaly.  This will be done in both the standard model (SM) and the minimal 
supersymmetric standard model (MSSM).
We will then set up our formalism for treating gravity theories with CCSFs in Section 3.
In Section 4, we will present the results for the cosmological evolution of the 
scale factor and scalar field.  A discussion and conclusions will be given in Section 5.

\section{Equation of state of the SM and the MSSM at high temperatures}
\label{thermo}

From Eq.~(\ref{eom}), the source term for the equation of motion
for a CCSF, is proportional to $\rho - 3p = \rho( 1- 3w)$ where the 
parameter $w$ characterizes the equation of state.  In the radiation dominated era, 
and in the absence of interactions, $w = 1/3$.
Here, we compute corrections to $w$ in the context of the SM as well as the MSSM at temperatures higher than the masses of the particles. This computation is readily extended to scalar-tensor theories of gravity, since they satisfy the weak equivalence principle (namely, non-gravitational physics is unaffected by the scalar field). This is manifest in the Jordan frame, in which matter is not directly coupled to the scalar. However, it is also easy to interpret the computation in the Einstein frame [see the discussion after Eq.~(\ref{rho-p-frames})]. It is useful to start from the expression for the free energy density, which we parametrize as
\begin{eqnarray}
{\cal F}
&=&-g_f\,\frac{\pi^2}{90}\,T^4\,.
\label{free}
\end{eqnarray}
In this expression, $g_f$ is the {\it effective free energy number of relativistic degrees of freedom}, 
that is defined by normalizing to the free energy density of one non-interacting massless 
bosonic degrees of freedom:
\begin{eqnarray}
{\cal F}_{1,{\rm free}} &=&-\frac{\pi^2 T^4}{90}\,.
\label{free-0}
\end{eqnarray}

For vanishing chemical potential, the pressure and energy density can be obtained from Eq.~(\ref{free}) as
\begin{eqnarray}
P&=&- {\cal F} \,, \nonumber\\
\rho  &=& T \, \frac{\partial P}{\partial T} - P = \frac{\pi^2}{90} T^4 
\left(3\,g_f+ \frac{\partial g_f}{\partial \, {\rm ln } \, T}
\right)\,.
\label{p-rho}
\end{eqnarray}
From these expressions, one finds
\begin{eqnarray}
\rho - 3P = \frac{\pi^2}{90} T^4\, \frac{\partial g_f}{\partial \, {\rm ln } \, T} \,,
\label{trace}
\end{eqnarray}
or, equivalently,
\begin{equation}
w \equiv \frac{P}{\rho} = \left( 3 + \frac{\partial \, {{\rm ln} \, g_f}}{\partial \, {\rm ln } \, T} \right)^{-1}\,.
\end{equation}

We see that the trace of the energy-momentum tensor (\ref{trace}) is different from zero whenever the effective number of degrees of freedom changes with the temperature. Most early universe calculations in cosmology consider only the contribution of the relativistic species to the pressure and energy density (since those of massive species are exponentially suppressed). Therefore, one includes only particles with mass smaller than the temperature in the expressions (\ref{p-rho}), and, as a consequence, $g_f$ only varies whenever the temperature drops below any mass threshold. However, $g_f$ also varies at temperatures above all mass thresholds in the theory, due to the trace anomaly \cite{brax}. To compute this, we recall the relation
\begin{equation}
{\cal F} = - \frac{T}{V} \, {\rm ln } \, Z
\end{equation}
where $Z$ is the partition function and $V$ the physical volume (one typically imposes periodic boundary conditions on a cube of total volume $V$; assuming homogeneity on a sufficiently large volume, the pressure and energy density do not depend on $V$). The partition function can then be computed diagrammatically \cite{Kapusta:1989tk}. The one loop vacuum diagram provides the free theory result (\ref{free-0}). Once we factor out the numerical factor $- \pi^2 / 90$, the coefficient $g_f$ is simply the number of bosonic and fermionic relativistic degrees of freedom
\begin{eqnarray}
g_{f,{\rm free}} &=& N_b+\frac{7}{8}\; N_f\,,
\end{eqnarray}
where the $7/8$ coefficient multiplying the fermionic contribution is due to the different value of the partition function obtained from the Fermi-Dirac thermal distribution rather than the Bose-Einstein distribution. In the SM, we have $N_b=28$ bosons and $N_f=90$ fermions, giving $g_{f,{\rm free}}  = 427/4 = 106.75 \,$.

\begin{figure}[h]
\centerline{
\includegraphics[width=0.4\textwidth,angle=0]{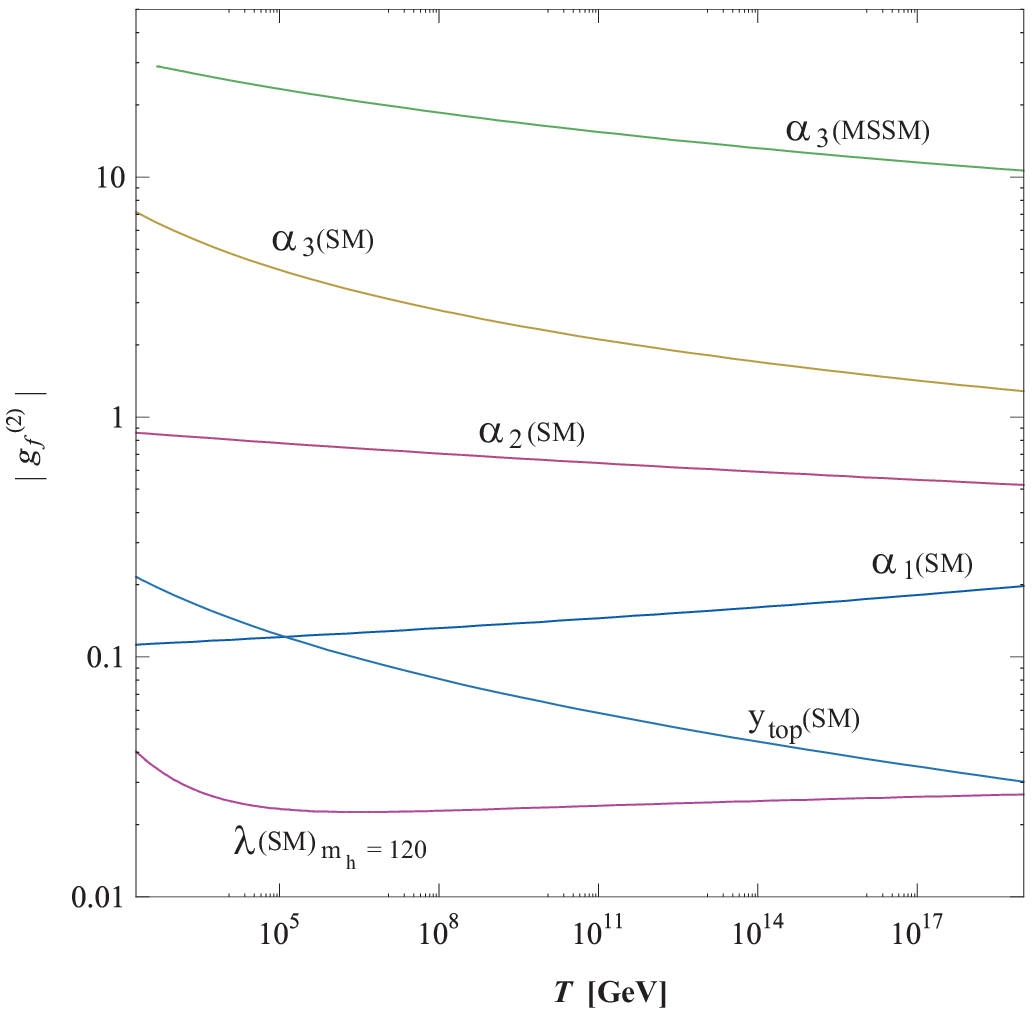}
\hspace{1cm}
\includegraphics[width=0.4\textwidth,angle=0]{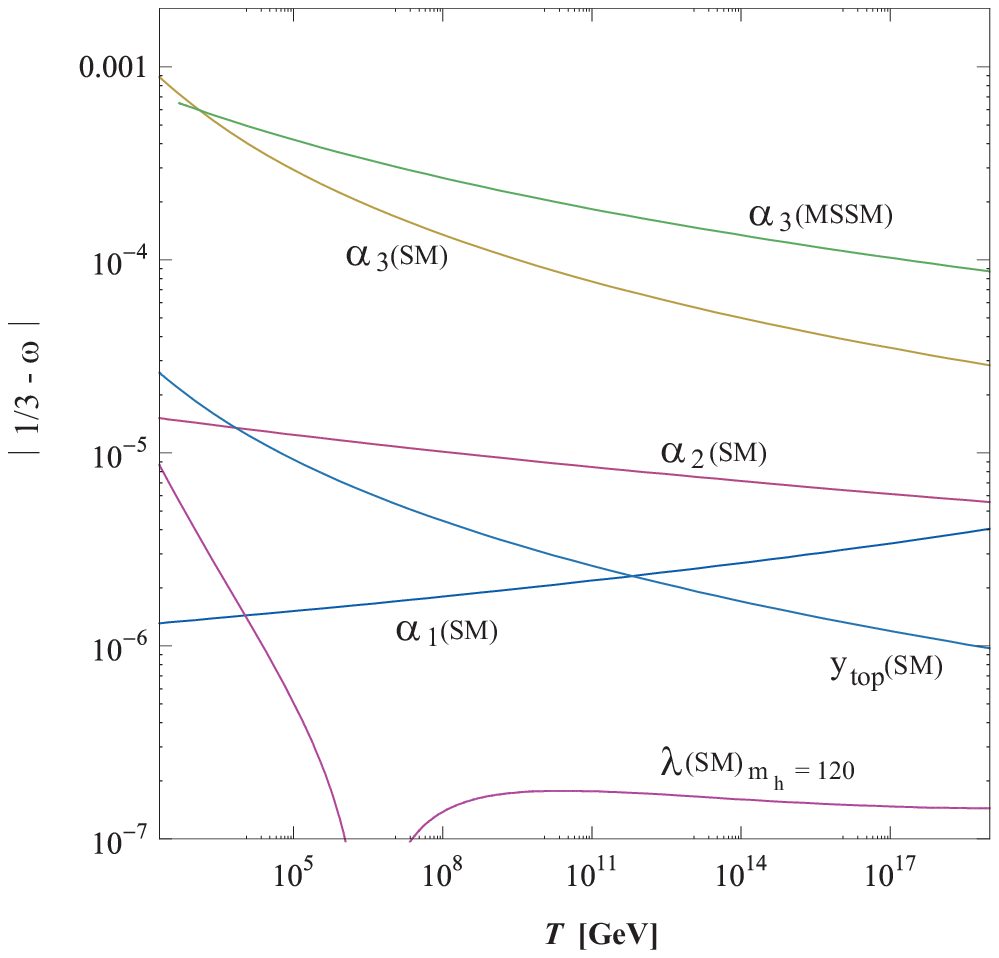}
}
\caption{Left panel: The absolute value of the leading corrections (two-loops) to the effective free energy number of relativistic
degrees of freedom from the different gauge interactions, top Yukawa and Higgs self-coupling in the SM.  The strong contribution 
($\alpha_3$), is dominant and is also plotted for the case of the MSSM assuming that all supersymmetric partners
have masses fixed at $m_{\su} = 500 \; \gev$. All contributions to $g_{f}^{(2)}$ 
are negative.
Right panel: The absolute value of the leading correction to $1/3-\omega$ from the same interactions shown in the left panel. The strong 
contribution is dominant  not only due to the fact that the correction to $g_{f}^{(2)}$ is bigger, but also because 
the running of $\alpha_3$ is steeper. On the other hand, the contribution from the hypercharge interaction (and the Higgs coupling 
for high temperatures) is negative and opposite to the others since $\alpha_1$ ($\lambda$) increases with $T$.
}
\label{fig:w-sm-mssm}
\end{figure}

Higher loops account for the interactions. The departure from the ideal gas regime due to the interactions reduces the effective number of degrees of freedom. Here, we only consider the leading order correction encoded in two loop vacuum diagrams.\footnote{As we shall see, the effect is dominated by the strong interactions; at temperatures just above the masses of the SM particles, higher order loops are also important. The various loop contributions for QCD, up to ${\alpha_3}^3 \ln(1/\alpha_3)$ have been evaluated in Ref.~\cite{Kajantie:2002wa}, where it is shown that only for temperatures $T>10^5\gev$, the two loop contribution dominates higher order terms. As we are interested in the dynamics of the Universe at very high energies, we will discuss only the leading term and assume that is a good approximation.} The corrections are proportional to the gauge, Yukawa, and Higgs couplings. The variations of these couplings with temperature then lead to the nontrivial dependence of $g_f$ on the temperature, and, as a consequence, to $\rho - 3 \, P \neq 0 \,$ (see Fig. \ref{fig:w-sm-mssm}).

For the SM, the one and two loop contributions give
\cite{Kapusta:1989tk,Kajantie:2002wa,Andersen:1997hq,arnold1,kast,braaten2,parw1}:
\begin{eqnarray}
\label{free-SM}
g_f
&=&\frac{427}{4}
-\frac{210}{\pi}\,\alpha_3
-\frac{645}{8\pi}\,\alpha_2
-\frac{165}{8\pi}\,\alpha_1
\\
& &-\frac{15}{8\,\pi^2}\,\lambda
-\frac{25}{32\,\pi^2}
\,
\left(
3|y_{\alpha\beta}^{U}|^2
+3|y_{\alpha\beta}^{D}|^2
+|y_{\alpha\beta}^{L}|^2
\right)\,,
\nonumber
\end{eqnarray}
where
\begin{eqnarray}
|y_{\alpha\beta}^{I}|^2
&=&
y_{\alpha\beta}^{I}{y^{I}}^\dagger_{\beta\alpha}
\,,
\nonumber
\end{eqnarray}
are the Yukawa matrices of the SM; $\lambda$ is the quartic Higgs coupling, and $\alpha_3$, 
$\alpha_2$, $\alpha_Y$ are the coupling constants associated with the strong, weak and hypercharge 
interactions respectively. Our results are expressed in terms of the coupling 
$\alpha_1=5\alpha_Y/3$, 
(we are following notation from 
Ref.~\cite{Arason:1991ic}).
Explicitly:
\begin{eqnarray}
\alpha_1= \frac{5g_Y^{2}}{12\pi}=\frac{5\alpha}{3\cos^2\theta_W}
\;\;\;,\;\;\;
\alpha_2= \frac{g_2^2}{4\pi}=\frac{\alpha}{\sin^2\theta_W} 
\;\;\;,\;\;\;
\alpha_3= \frac{g_3^2}{4\pi},
\end{eqnarray}
where $\alpha=e^2/4\pi$, and $g_Y$, $g_2$ and $g_3$ are the usual $U_Y(1)$, $SU(2)$ and $SU(3)$
coupling constants. We are using the modified minimal subtraction ($\overline{MS}$) 
scheme~\cite{msbar} as a renormalization prescription.

Results from LEP and the Tevatron imply~\cite{Amsler:2008zzb}:
\begin{equation} 
\alpha_1(M_Z)\simeq 0.017 ,\ \ 
\alpha_2(M_Z)\simeq 0.034 , \ \ 
\alpha_3(M_Z)\simeq 0.118.
\end{equation}
%
At unification scales ($T \sim 10^{16} \gev$), the electroweak couplings are comparable to 
the strong coupling and their contributions could be more important. Eq. (\ref{free-SM}) and 
Fig. \ref{fig:w-sm-mssm} show, however, that this is not the case, and the 
strong contribution ($\alpha_3$) is dominant even at very high energies due to the 
large number of
strongly interacting particles. In addition, the third generation dominates the contribution of 
the Yukawa couplings. By taking into account the SM masses~\cite{Amsler:2008zzb,Group:2009ec}: 
$m_t\simeq 173~\gev$, $m_b\simeq 4.2~\gev$ and $m_\tau\simeq 1.78~\gev$, it is evident that we can 
neglect all the Yukawa contributions apart from the top:
\begin{eqnarray}
|y_{\alpha\beta}^{U}(m_t)|^2\simeq 2 \frac{{m_t}^2}{v^2}\simeq\,0.99\,, 
\end{eqnarray}
where the vacuum expectation value of the Higgs is $v\simeq 246~\gev$. Finally, $\lambda$ 
is the only unknown parameter of the SM. In the simplest model, It is related directly to the Higgs mass: $\lambda=(m_h/v)^2$. 
Combined results from the four LEP collaborations (ALEPH, DELPHI, L3 and OPAL)
gives the lower bound \cite{Amsler:2008zzb}: $m_h > 114.4~\gev$  at $95\%$ of confident level,
which implies 
\begin{eqnarray}
\lambda(m_h) > 0.216\,.
\end{eqnarray}
In Fig. \ref{fig:w-sm-mssm}, we have fixed $m_h = 120~\gev$.

As mentioned earlier, the temperature dependence of the couplings leads to the deviation from an ideal 
relativistic gas. This dependence is described by the renormalization group equations (RGE). Taking the 
leading  (one-loop) contribution \cite{Arason:1991ic} to the running, we find:
\begin{eqnarray}
\frac{\partial g_f}{\partial \, {\rm ln } \, T}
&=& \frac{1}{80 \, \pi^2} \left( - 3423 \, \alpha_1^2 - 135 \, \alpha_1 \, \alpha_2 + 9875 \, \alpha_2^2 + 58800 \, \alpha_3^2 \right) \nonumber\\
&&+ \frac{15}{256 \, \pi^3} \left( 17 \, \alpha_1 + 45 \, \alpha_2 + 160 \, \alpha_3 + \frac{3}{2 \, \pi} \, y_t^2 \right) y_t^2 \nonumber\\
&& + \frac{9 \lambda}{32 \pi^3} \left( 3 \, \alpha_1 + 15 \, \alpha_2 - \frac{5}{\pi} y_t^2 - 5 \, \lambda \right) 
\,.
\label{dgf-sm}
\end{eqnarray}
The contribution to $\partial g_f / \partial {\rm ln } \, T$ from the strong interaction is even more dominant 
than the contribution to $g_f$, since the running of $\alpha_3$ is also more important than that of the other couplings. In this case, keeping only the QCD interaction is a good approximation up to order $10\%$, over the entire range of energies considered, unless the Higgs is very heavy ($m_h\gta 170\; \gev$).
 At high energies the quartic Higgs interaction (and the hypercharge coupling) runs in the opposite direction to that of $\alpha_3$, which partially compensates for the QCD contribution. 

The greatest source of uncertainty in this computation is the effect of higher order QCD terms in Eq. (\ref{free-SM}). To estimate this effect, we included terms up to ${\mathcal O } \left( \alpha_3^{5/2} \right)$ in the computation of $g_f$ for the SM (the various terms are listed in Ref.~\cite{Kajantie:2002wa}), and computed the corresponding value for $w$. We found a value for $1/3 - w$ about $50\%$ smaller than that obtained from the ${\mathcal O } \left( \alpha_3 \right)$ term alone reported in (\ref{free-SM}), in the entire range of temperatures $10 \, {\rm GeV} < T < 10^{15} \, {\rm GeV}$. Unfortunately, the next order contribution has an undetermined coefficient, and it has been shown in Ref.~\cite{Kajantie:2002wa} that this could have a non negligible effect on $g_f$. Moreover, there are nonperturbative contributions which could also be substantial. Due to these uncertainties, we choose to retain only the first nontrivial, ${\mathcal O} \left( \alpha_3 \right)$, contribution to $g_f$,``comforted'' by the fact that, summing up to the highest fully known order, gives a result which is in reasonable agreement with the one we adopt.

In summary, for $m_h\lta 170\,\gev$, we have the approximate result
\begin{eqnarray}
&&g_f = \frac{427}{4} - 420 \, {\tilde \alpha}_3 + {\mathcal O } \left( {\tilde \alpha}_3^{3/2} \right)
\;\;\;,\;\;\; \frac{\partial g_f}{\partial \, {\rm ln } \, T} 
= 2940 \, {\tilde \alpha_3}^2 + {\mathcal O } \left( {\tilde \alpha}_3^{5/2} \right), \nonumber\\
&& w - \frac{1}{3} = - \frac{560}{183} \, {\tilde \alpha}_3^2 + {\mathcal O } \left( {\tilde \alpha}_3^{5/2} \right)
\;\;\;,\;\;\;
{\tilde \alpha}_3 \left(T\right) \equiv \frac{\alpha_3}{2 \pi} \simeq\frac{{\tilde \alpha}_3 \left( m_t \right)}{1+7 \, {\tilde \alpha}_3 \left( m_t \right) \, \ln\left( \frac{T}{m_t}\right)},
\label{SM}
\end{eqnarray}
with ${\tilde \alpha}_3 \left( m_t \right)\simeq 0.0172$. 

We can repeat this computation for the MSSM. Including only the (super)-QCD contribution, we find
\begin{eqnarray}
&&g_f = \frac{915}{4} - 1890 \, {\tilde \alpha}_3 + {\mathcal O } \left( {\tilde \alpha}_3^{3/2} \right)
\;\;\;,\;\;\; \frac{\partial g_f}{\partial \, {\rm ln } \, T} 
= 5670 \, {\tilde \alpha_3}^2 + {\mathcal O } \left( {\tilde \alpha}_3^{5/2} \right), \nonumber\\
&& w - \frac{1}{3} = - \frac{168 \, {\tilde \alpha}_3^2}{61} + {\mathcal O } \left( {\tilde \alpha}_3^{5/2} \right) 
\;\;\;,\;\;\;  
{\tilde \alpha}_3 \left(T\right)  \simeq\frac{{\tilde \alpha}_3 \left( m_\su \right)}{1+3 \, {\tilde \alpha}_3 \left( m_\su \right) \, \ln\left( \frac{T}{m_\su}\right)},
\label{MSSM}
\end{eqnarray}
Assuming for definiteness that the supersymmetric partners have masses $m_{\su} = 500 \, {\rm GeV}$, and using the running in Eq.~(\ref{SM}) from $T=m_Z$ to $T=m_\su$, we have $\alpha_3(m_{\su})\simeq 0.0153$.

We recall that the expressions (\ref{SM}) and (\ref{MSSM}) hold for temperatures greater than the masses of the particles, i.e. for $T \gta 200 \, {\rm GeV}$ in the SM, and (with our assumption) for $T \gta 500 \, {\rm GeV}$ in the MSSM. In the right panel of Figure \ref{fig:w-sm-mssm} we show different contributions to the departure of the equation of state from the value $w = 1/3$ value 
corresponding to the noninteracting case. We note that the dominant effect decreases as the temperature increases, since  $\alpha_3$ decreases with temperature (and, consequently the absolute value of the two loop contribution to $g_f$ decreases). We also notice that, at the lowest temperatures shown, the departure is stronger in the SM than in the MSSM case. This is because $\alpha_3$ runs faster in the SM .
However, since $\alpha_3$ decreases less in the MSSM, the departure from the noninteracting value becomes greater for the MSSM as the temperature increases.

\section{Scalar-tensor theories of gravity}

\label{cosmology-eqs}

In scalar-tensor  theories, gravity is mediated by both a spin-$2$ graviton and a spin-$0$ scalar field that couples universally to matter. Following the notation of Refs.~\cite{couv,ru02,carlo}, we start from the action
\begin{equation}
S = \int \frac{d^4 x}{16 \pi G_*} \sqrt{-g} \left[ \frac{R}{A^2 \left( \phi \right)} - g^{\mu \nu} \partial_\mu \phi \, 
\partial_\nu \phi - 2 U \left( \phi \right) \right] + S_m \left[ g_{\mu \nu} \right]
\label{act-jordan}
\end{equation}
where the first term is the action for the spin-2 graviton and the scalar field, while the second term is the action for matter. This expression defines the theory in the Jordan frame, in which the standard expression of the metric $g_{\mu \nu}$ is used in the action for the matter 
fields (where by ``matter'' we generally denote any field in the theory apart from $\phi$
and $g_{\mu\nu}$). However, the gravitational interaction between matter fields is modified, since the function $A \left( \phi \right)$ multiplies the Ricci scalar. Such a theory depends on 2 arbitrary
functions and we shall  assume in this work a vanishing potential for the scalar field, $U = 0 \,$.
We emphasize that $G_*$ is the bare gravitational constant and does not correspond
to the gravitational constant that would be measured in a Cavendish-type of
experiment\footnote{It can be shown that Newton's constant is given by
$G_N=G_*A^2(\phi_*)(1+\alpha^2)$ which depends a priori on time through $\phi_*$.
The deviations from general relativity can be evaluated in terms
of the post-Newtonian parameters
$$
\gamma-1 = -2\frac{\alpha^2}{1+\alpha^2}\qquad
\beta - 1 = \frac{1}{2}\frac{\alpha^2}{(1+\alpha^2)^2}\alpha'
$$
as long as the field is light enough, as considered in this work; (see Ref.~\cite{gefp}).
\label{foot2}}.

Under a conformal transformation
\begin{equation}
g_{*\mu\nu} = A^{-2} \left( \phi \right) \, g_{\mu \nu}\,,
\end{equation}
and for a vanishing potential, the action of the system becomes
\begin{equation}
S = \int \frac{d^4 x}{16 \pi G_*} \sqrt{-g_*} \left[ R_* - 2 g_*^{\mu \nu} \partial_\mu \phi_* \partial_\nu \phi_*  \right] + S_m \left[ A^2 \left( \phi_* \right) g_{ *  \mu \nu} \right]
\label{act-einstein}
\end{equation}
where the scalar field $\phi_*$ is defined by
\begin{equation}
\left( \frac{d \phi_*}{d \phi} \right)^2 = 3 \left( \frac{d \, {\rm ln } \, A \left( \phi \right)}{d \phi} \right)^2 + \frac{A^2 \left( \phi \right)}{2}
\end{equation}
and $A \left( \phi_* \right)$ is short  for $A \left( \phi \left( \phi_* \right) \right) \,$. 

The expression (\ref{act-einstein}) is the action of the system in the Einstein frame. It is characterized by a standard action for the spin-2 graviton; however, the combination $A^2 \left( \phi_* \right) g_{ *  \mu \nu}$, rather than the metric itself, is used in the action for matter. While the two expressions are equivalent, the Einstein frame is more often used to study the cosmology of the system (since the resulting cosmological equations are the standard ones), while the Jordan frame is more often used to study particle physics processes (since the physical lengths and masses are constant in this frame). Moreover, the Einstein frame is built in such a way that the kinetic term of the spin-2 and spin-0 mediators is diagonal so that
the Cauchy problem is well-paused in this frame.

We consider a FLRW universe with Euclidean spatial sections, in the Einstein frame, the line element is
\begin{equation}
d s^2 = - d t_*^2 + R_*^2 \left( t \right) dx^i \, d x^i
\end{equation}
and we denote by $\rho_*$ and $P_*$, respectively,  the total energy density and pressure of the matter fields. We then find the evolution equations \cite{couv},
\begin{eqnarray}
&& 3 H_*^2 = 8 \pi G_* \rho_* + \left( \frac{d \phi_*}{d t_*} \right)^2 \nonumber\\
&& - \frac{3}{R_*} \frac{d^2 R_*}{d t_*^2} = 4 \pi G_* \left( \rho_* + 3 P_* \right) + 2 \left( \frac{d \phi_*}{d t_*} \right)^2 \nonumber\\
&& \frac{d^2 \phi_*}{d t_*^2} + 3 H_* \frac{d \phi_*}{d t_*} = - 4 \pi G_* \alpha \left( \phi_* \right) \left( \rho_* - 3 P_* \right) \nonumber\\
&&\frac{d \rho_*}{d t_*} + 3 H_* \left( \rho_* + P_* \right) = \alpha \left( \phi_* \right) \left( \rho_* - 3 P_* \right) \frac{d \phi_*}{d t_*}
\label{eq-einstein}
\end{eqnarray}
where the Hubble rate is defined as $H_* \equiv d\ln R_* / d t_* $, and $\alpha \left( \phi_* \right)$ was defined in Eq.~(\ref{alpha}). The first Eq. of (\ref{eq-einstein}) is the $00$ Einstein equation while the second equation is a linear combination of the $00$ and $ii$ Einstein equations. We note that these equations are the standard FLRW equations, and the scalar field contributes to them only through its energy density and pressure. The other two equations are the field equations for the scalar field, and for matter, respectively. One out of the last three equations can be derived from the others in Eq. (\ref{eq-einstein}), as a consequence of a nontrivial Bianchi identity. 
Both source terms are proportional to $\alpha$ and reflect
the scalar interaction and the action-reaction law (or the fact that the sum of the two stress-energy tensors is covariantly conserved). In the third equation, we see that the scalar field is coupled to the trace of the energy momentum tensor of matter as described in the introduction. This term is usually assumed to vanish in the radiation dominated era. However, as we discussed in the previous Section, this term is actually always nonvanishing at some level. Finally, we see from the last expression in Eq. (\ref{eq-einstein}) that the scalar field also modifies the continuity equation for matter. 

The geometry in the Jordan frame is also of the FLRW type, with scale factor and time related to those in the Einstein frame by
\begin{equation}
R = A \left( \phi_* \right) R_* \;\;\;,\;\;\; d t = A \left( \phi_* \right) d t_*\,.
\label{Rt-frames}
\end{equation}
As a ``time variable'' for the evolution, we choose the e-folds of expansion $p$ in the Einstein frame (not to be confused with the pressure $P$). Denoting by $R_{*,{\rm in}}$ the value of the scale factor at some initial time (for instance, at the end of reheating, when the thermal bath has formed), the number of e-folds is
\begin{equation}
p \equiv {\rm ln } \frac{R_*}{R_{*,{\rm in}}}\,.
\label{p-def}
\end{equation}
From the relations (\ref{Rt-frames}) and (\ref{p-def}), one can immediately 
relate the derivatives with respect to the time in either frame to derivatives 
with respect to $p$:
\begin{equation}
\frac{d}{d p} = \frac{1}{H_*} \, \frac{d}{d t_*} = \frac{A}{H_*} \, \frac{d }{d t}\,.
\label{chain}
\end{equation}
In what follows, prime will denote differentiation with respect to $p \,$.

It is worth noting that $R_*$ grows monotonically with time, since $H_*$ is governed by the standard Friedmann equation (the first of Eq.(\ref{eq-einstein})) and it is strictly positive, hence ensuring $H_*$ does not change sign
so that $p(t_*)$ is a monotonic function. Therefore, we can indeed use $p$ as ``time variable''. On the contrary, the Hubble rate $H$ in the Jordan frame is related to that of the Einstein frame by
\begin{equation}
H = \frac{H_*}{A} \left( 1 + \alpha \, \phi_*' \right)\,.
\label{hubbles}
\end{equation}
As we shall see in the next Section, it is possible that $H$ becomes negative, indicating that the universe is actually contracting in the Jordan frame for some time during the evolution. Therefore, the scale factor and the temperature in the Jordan frame can have a non monotonic evolution, and we cannot use either as ``time variables'' for the system.

The evolution equations in the Jordan frame are given in Ref.~\cite{couv} and we will not repeat them here. We only note that the continuity equation takes its standard form 
\begin{equation}
\frac{d \rho}{d t} + 3 \, H \, \left( \rho + P \right) = 0
\label{cont-j}
\end{equation}
in terms of the Hubble parameter in the Jordan frame, $H \equiv d\ln R / d t$. This is obvious since in the Jordan frame, the matter content is not directly coupled to $\phi$. On the other hand:
\begin{equation}
\rho \equiv \rho_* / A^4 \;,\;\;\;\;\; P \equiv P_* / A^4\;.
\label{rho-p-frames}
\end{equation}
These expressions relate  the energy density and pressure in the Jordan frame, to those in the Einstein frame.

To proceed, we need to express the pressure and energy density in terms of the temperature, through the relations obtained in the previous Section. We can do this in either frame. Indeed, the pressure and energy densities, as well as the free energy, can be written as $T^4$ times a function of $T/m$  (for example, $m_t$ in the SM , Eq.~(\ref{SM}), or $m_\su$ in the MSSM, Eq.~(\ref{MSSM})). From Eq.~(\ref{rho-p-frames}) we deduce that the temperatures $T$ in the Jordan frame and $T_*$ in the Einstein frame are related to each either by $T = T_* / A \,$. An analogous relation takes place between the mass of a particle in the two frames, $m = m_* / A \,$.~\footnote{To see this, consider the  action of a massive field in the Jordan frame (where it is of the standard form, and the mass parameter is $m$), transform to the Einstein frame, rescale the field so that it is canonically normalized in this frame, and observe that the mass parameter of the newly canonically normalized field has indeed become $A \, m \equiv m_*$.} Therefore, the ratio between the temperature and the mass scale is the same in both frames, and we can use relations like (\ref{SM}) and (\ref{MSSM}) in either frame, with the rescaling of $T$ and $m_t$ (or $m_\su$). 

For definiteness, we use these relations in the Jordan frame, where the mass scales are constant. 
Inserting the expressions (\ref{p-rho}) for the energy and pressure in the continuity equation (\ref{cont-j}), we get
\begin{equation}
\frac{4}{T} \, \frac{d T}{d t} \left(3\,g_f+ \frac{\partial g_f}{\partial \, {\rm ln } \, T} \right) + 
\frac{d }{d t} \left(3\,g_f+ \frac{\partial g_f}{\partial \, {\rm ln } \, T} \right) + 
\frac{3 }{R} \, \frac{d R}{d t} \left(4\,g_f+ \frac{\partial g_f}{\partial \, {\rm ln } \, T} \right)
= 0\,.
\label{TR-par}
\end{equation}
Since $g_f$ is a function of the temperature,
\begin{equation}
\frac{d g_f}{d t} = \frac{\partial g_f}{\partial \, {\rm ln } \, T} \, \frac{1}{T} \, \frac{d T}{d t}\,.
\end{equation}
Inserting this into Eq.~(\ref{TR-par}), we obtain
\begin{equation}
\frac{d \, {\rm ln } \, T}{d \, t}  = -  \left( \frac{12 g_f + 3 \, \frac{d g_f}{d \, {\rm ln } T}} 
{12 g_f + 7 \, \frac{d g_f}{d {\rm ln } T} + \left( \frac{d}{d {\rm ln } T} \right)^2 g_f} \right) 
\frac{d  \, {\rm ln } \, R}{d \, t}
\equiv - \frac{1 }{\cal I} \, \frac{d \, {\rm ln } \, R}{d \, t}\,.
\label{TR}
\end{equation}
Using equations (\ref{SM}) and (\ref{MSSM}) we have
\begin{eqnarray}
{\cal I} &=& 1+ \frac{560}{61} \, {\tilde \alpha}_3^2 + {\mathcal O} \left( {\tilde \alpha}_3^{5/2} \right) \;\;\;,\;\;\; {\rm SM} \nonumber\\
{\cal I} &=& 1+ \frac{504}{61} \, {\tilde \alpha}_3^2 + {\mathcal O} \left( {\tilde \alpha}_3^{5/2} \right) \;\;\;,\;\;\; {\rm MSSM}. \nonumber
\end{eqnarray}
In both cases, ${\cal I} -1 \simeq 0.002$ at the electroweak scale, and it decreases further as the temperature grows (for instance, ${\cal I} -1 = {\mathcal O} \left( 10^{-4} \right)$  at $T = 10^{16} \, {\rm GeV} \,$). If we ignore this small effect, and simply set ${\cal I} \equiv 1 \,$, Eq. (\ref{TR}) gives us
the standard relation $T \propto 1 / R \,$ as long as there are no other processes involved, such particle annihilations which can reheat the thermal bath. Since $T_* = A \, T$, and $R_* = R / A$, the same inverse proportionality $T_* \propto 1 / R_*$ also holds in the Einstein frame.

We can also replace the time derivative on the left hand side of Eq.~(\ref{TR}) with a derivative 
with respect to $p$ through Eq. (\ref{chain}). Recalling that $d \, {\rm ln } \, R / d \, t = H$, and the relation (\ref{hubbles}) between the Hubble parameters in the two frames, we find
\begin{equation}
\left( {\rm ln } \, T \right)' = - \frac{1+ \alpha \, \phi_*'}{\cal I}
\label{finaleqT}
\end{equation}
where we recall that prime denotes differentiation with respect to $p\,$.

Since $\alpha = d \, {\rm ln } \, A / d \, \phi_*$, if we again neglect the small departure of ${\cal I}$ from one, this relation can be integrated to give
\begin{equation}
\frac{T}{T_{\rm in}} = \frac{R_{\rm in}}{R} = \frac{A \left( \phi_{*,{\rm in}} \right)}{A \left( \phi_* \right)} \, {\rm e}^{-p} \;\;\;,\;\;\; {\rm for  } \;\; {\cal I} \equiv 1
\label{solT}
\end{equation}
where we recall that, by definition, $p_{\rm in} = 0 \,$. We also have
\begin{equation}
\frac{T_*}{T_{*,{\rm in}}} = \frac{R_{*,{\rm in}}}{R_*} = {\rm e}^{-p} \;\;\;,\;\;\; {\rm for  } \;\; {\cal I} \equiv 1.
\end{equation}

The cosmology of the system is therefore completely specified once the evolution $\phi_* \left( p \right)$ is known. Starting from the third expression of (\ref{eq-einstein}), and replacing the time derivatives with derivatives with respect to $p$ through (\ref{chain}), we arrive at
\begin{equation}
\phi_*'' = - \frac{3 - \phi_*^{' 2}}{2} \left[ \left( 1 - w \right) \phi_*' + \left( 1 - 3 w \right) \alpha \right]
\nonumber\\
\label{finaleqp}
\end{equation}
where the equations of state $w$ for the MS and for the MSSM is given in Eqs. (\ref{SM}) and (\ref{MSSM}), respectively.\footnote{These are good approximations as long as we do not cross any mass threshold.}
The equation of state is completely determined by the temperature $T$. Therefore, we can either solve the two equations (\ref{finaleqT}), and (\ref{finaleqp}) numerically, or we can neglect the small departure of ${\cal I}$ from unity, and use the analytic 
expression (\ref{solT}) for the temperature. Our results below are based on the numerical solution, however, we have verified that $T \, R = T_* \, R_*$ remains constant with an extremely good accuracy.

\section{Cosmological solutions}

\label{cosmology-sols}

There are strong limits today on deviations from general relativity~\cite{Will:2005va}. The bounds are most conveniently imposed on the so called post-Newtonian parameters (see Footnote~\ref{foot2}), which in the context of scalar-tensor theories of gravity, can in turn be related to the present values $\alpha_0$ and $\beta_0$ of the function (\ref{alpha}), and its derivative 
\begin{equation}
\beta \left( \phi_* \right) \equiv \frac{d \alpha}{d \phi_*}
\label{defbeta}
\end{equation}
Limits from the Very Long Baseline Interferometer \cite{Shapiro:2004zz} and the Cassini spacecraft 
\cite{Bertotti:2003rm} enforce $\alpha_0^2 \lta 10^{-5} \,$. The perihelion shift of Mercury \cite{mercurybound} and the Lunar Laser Ranging experiment  \cite{Williams:1995nq} instead constrain the combination $\vert (\beta_0 + 1) \, \alpha_0^2 \vert \lta {\mathcal O} \left( 10^{-3} \right)$. Therefore, $\beta_0$ can be quite large, provided that $\alpha_0$ is sufficiently small. As argued in Ref.~\cite{couv}, one should however assume that $\beta_0 \lta {\mathcal O} \left( 100 \right)$, so that the post-Newtonian approximation scheme makes sense. 

These stringent limits do not need to hold at earlier cosmological eras, and, in fact, for very large 
$\beta$,  $\alpha$ could have been large in the past as well. This is due to the attraction towards GR that this type of scalar-tensor theory of gravity possesses \cite{dnord}. This can be easily seen from the evolution equation (\ref{finaleqp}) for the scalar field. This is an equation for a relativistic particle, 
subject to friction and evolving in a potential
\begin{equation}
V_{\rm eff} = \frac{3}{2} \left( 1 - 3 w \right) \int \alpha \, d \phi_* = \frac{3}{2} \left( 1 - 3 w \right) {\rm ln } \, A \left( \phi_* \right)
\end{equation}
(up to an irrelevant constant). Following the conventions in the literature, we define
\begin{equation}
A \left( \phi_* \right) \equiv {\rm e}^{a \left( \phi_* \right)} \;\;\;\Rightarrow\;\;\; \alpha = \frac{d a}{d \phi_*} \;\;\;,\;\;\; \beta = \frac{d^2 a}{d \phi_*^2} \;\;,\;\; V_{\rm eff} = \frac{3}{2} \left( 1 - 3 w \right) a \left( \phi_* \right)\,.
\label{definitions}
\end{equation}
The evolution naturally tends toward a minimum of the effective potential, for which $\alpha = 0 \,$,
if such a minimum exists. This is more effective for large curvature, namely for large values of $\beta \,$. Clearly, we need to assume that $a$ has a minimum; this excludes the simplest scalar-tensor theory, namely the Brans-Dicke one, for which $a$ is linear in $\phi_*$. In our concrete examples below, we consider the simplest example which satisfies the attraction towards GR mechanism, namely the quadratic coupling function\footnote{In practice, we are Taylor expanding the function $a \left( \phi_* \right)$ around its minimum, assuming that higher order terms are irrelevant. A constant term in $a$ is irrelevant, since it simply changes the constant $G_*$ and the normalization of the scalar field 
[cf. the action~(\ref{act-jordan})]. A linear term can be reabsorbed with a shift of $\phi_* \,$. }
\begin{equation}
a \left( \phi_* \right) = \frac{\beta}{2} \, \phi_*^2\,.
\label{our-a}
\end{equation}

The force that drives $\alpha$ towards zero requires that matter fields are coupled to $\phi_*$ through a nonvanishing trace of the stress energy tensor (notice that, indeed, $V_{\rm eff}$ vanishes for $w=1/3$). The present literature disregards the effect of the trace anomaly pointed out in Section~\ref{thermo}, but studies the departure from $w=1/3$ taking place during the matter dominated era, and when the temperature crosses any mass threshold.

It is easy to use Eq. (\ref{finaleqp}) to estimate the decrease of $\alpha$ during matter domination \cite{dnord}. Assuming that the scalar field is sufficiently small so that $\phi_*' \ll \sqrt{3} \,$, \footnote{In this case, Eq. (\ref{finaleqp}) is a linear equation, and the ratio between the final and initial values of $\phi_*$ is independent of the initial value of $\phi_*$.} and that the difference between the Jordan and Einstein frames can be ignored in this estimate, Eq. (\ref{finaleqp}) is solved by a simple damped oscillatory solution. Looking only at the decrease of the amplitude of the oscillations, we find
\begin{equation}
\Big\vert \frac{\alpha_0}{\alpha_{\rm eq}} \Big\vert^2 = \Big\vert \frac{a_0}{a_{\rm eq}} \Big\vert  \simeq {\rm e}^{- \lambda \, \left( p_0 - p_{\rm eq} \right)} \simeq z_{\rm eq}^{-\lambda} \;\;\;,\;\;\; \lambda = {\rm Re } \left\{ - \frac{3}{2} \left[ 1 - \left( 1 - \frac{8}{3} \, \beta \right) \right]^{1/2} \right\}
\label{phi-sol-const}
\end{equation}
(the suffixes ``eq'' and ``0'' denote the values of a quantity at the matter-radiation equality era and today, respectively). For $\beta > 3/8$ (we assume this to be the case in this work), we find that $a$ decreases by a factor of about $5\times10^3$ during the matter dominated era. This estimate can be compared to Figure 11 in the first reference in  Ref.~\cite{couv}. 
 
The decrease when the temperature drops below some mass threshold is more complicated. Refs. \cite{dnord} and \cite{couv} studied this effect, in relation to bounds imposed on these theory from BBN. Fig. 5 of  Ref.~\cite{couv} shows the ``source term'' (the coefficient that multiplies $\beta \phi_*$ in the effective potential) due to the mass threshold for any SM particle. One finds one distinct peak corresponding to the electron mass threshold, and a series of overlapping peaks corresponding to the other mass thresholds, so that there are two distinct phases in which $\phi_*$ decreases. The actual decrease of $a$ is strongly sensitive to the values of $\beta$, and $\phi_*$, and we refer to  \cite{couv} for more details. It is possible that $a$ decreases of a factor of ${\mathcal O} \left( 10^{-4} - 10^{-3} \right)$ in either phase.

We now turn to the departure from $w = 1/3$ pointed out in Section~\ref{thermo}. We focus on the early cosmology before BBN, at temperatures higher than any mass threshold. Specifically, we consider temperatures greater than $200 \, {\rm GeV}$ for the SM, and greater than $500 \, {\rm GeV}$ for the MSSM (as we mentioned in Section~\ref{thermo}, we assume that this is the mass of all of the supersymmetric partners of the SM).\footnote{The numerical values refer to the temperature in the Jordan frame, since this is the frame in which the particles have constant mass; notice however that the ratio between the temperature and any mass scale is identical in the two frames, see the discussion after Eq. (\ref{rho-p-frames}).}

\begin{figure}[ht!]
\centerline{
\includegraphics[width=0.4\textwidth,angle=-90]{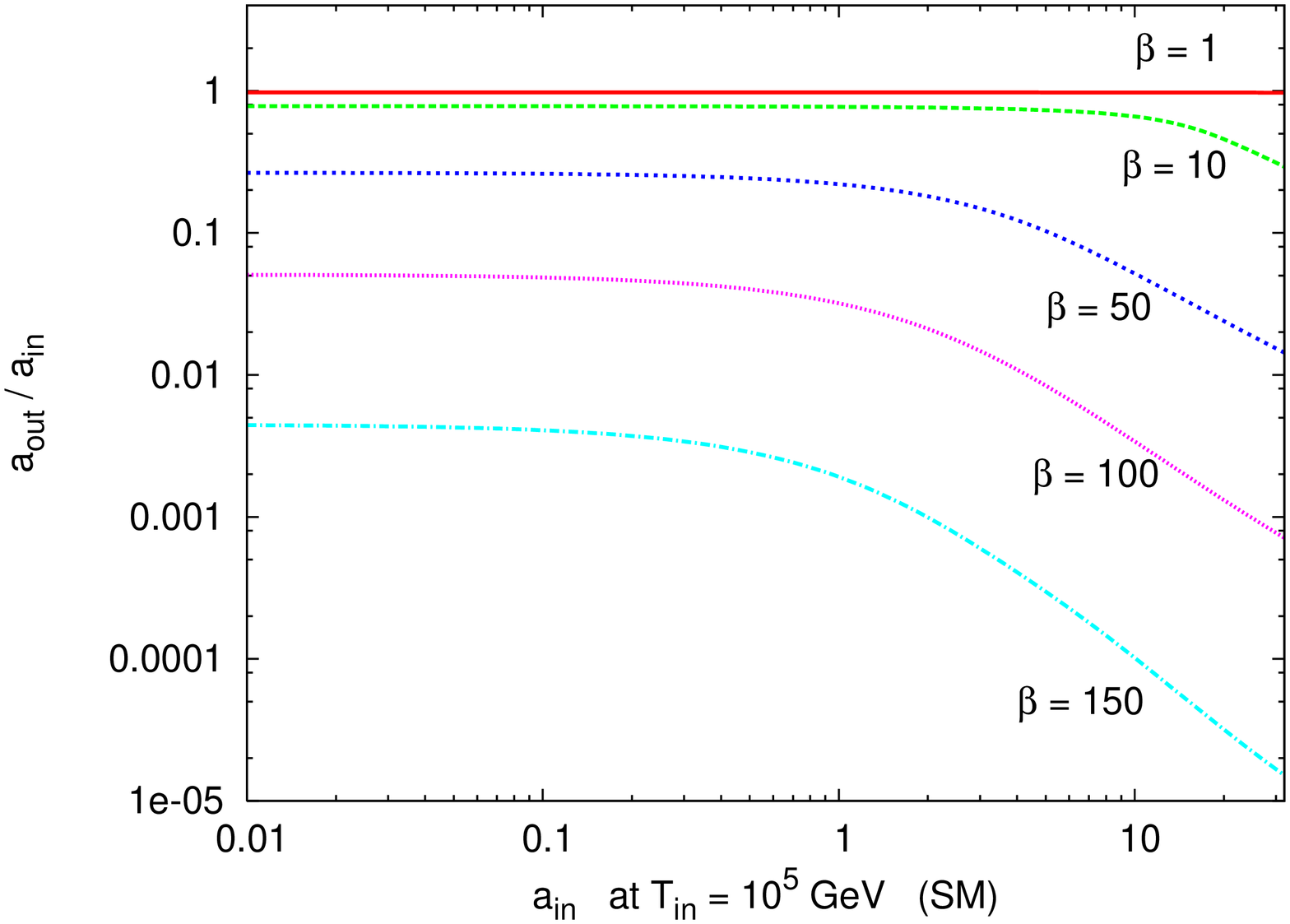}
\includegraphics[width=0.4\textwidth,angle=-90]{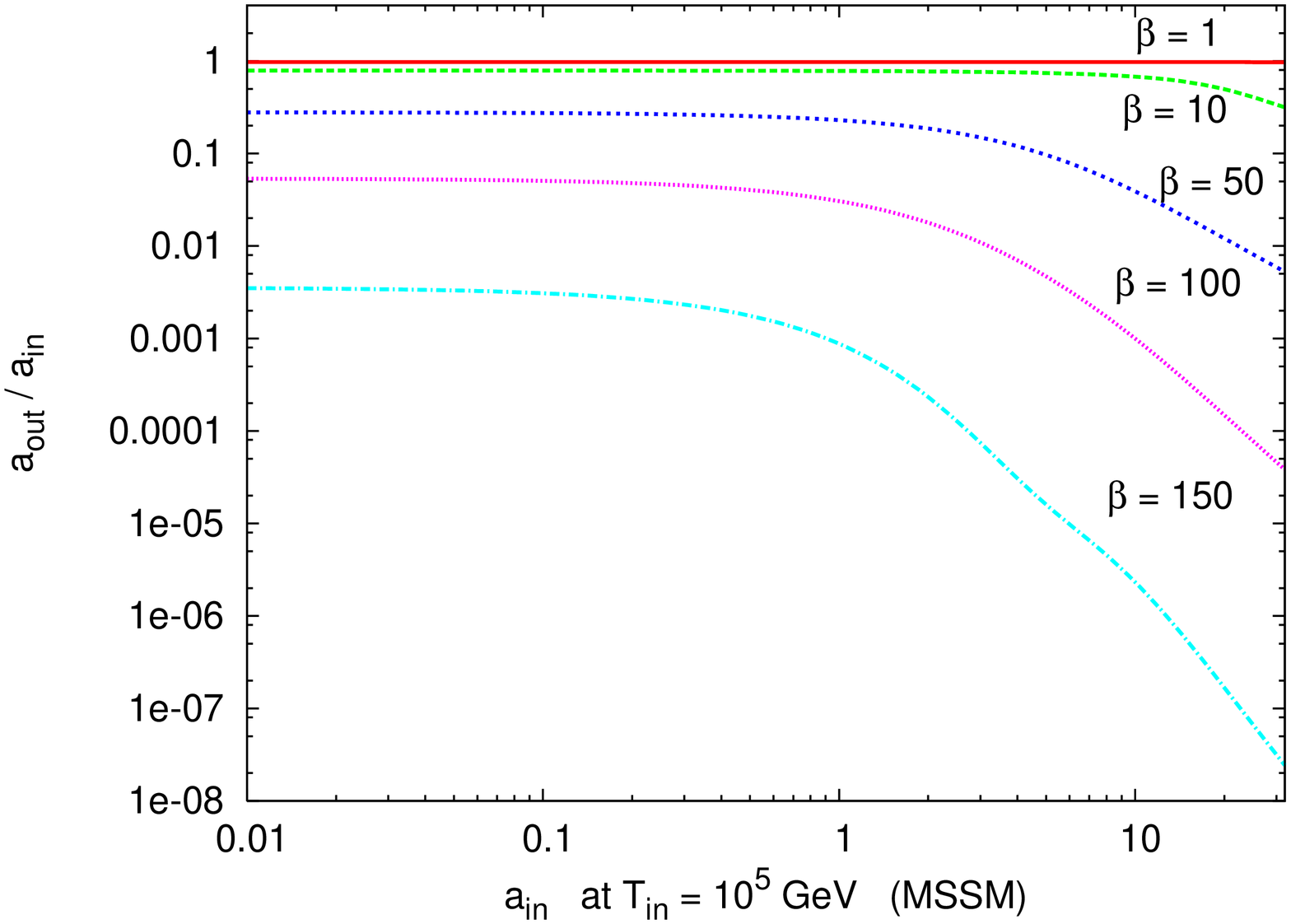}
}
\caption{Left panel: The decrease of $a = \beta \, \phi_*^2 / 2$ for the SM, starting from different values of $a$ at the initial temperature $T = 10^5 \, {\rm GeV}$ in the Jordan frame, and ending at the temperature $T =  200 \, {\rm GeV} \,$. Right panel: Decrease of $a$ in the MSSM with the same initial temperature, but with the final temperature $T = 500 \, {\rm GeV}$. See the main text for details.
}
\label{fig:sm-mssmt5}
\end{figure}

We show in Figure \ref{fig:sm-mssmt5} the decrease of $a \left( \phi_* \right)$ due to the trace anomaly. 
For illustrative purposes, we choose the initial temperature $T_{\rm in}= 10^5 \, {\rm GeV}$ in the Jordan frame. We then evolve the two equations (\ref{finaleqT}) and (\ref{finaleqp}) starting from several different values of $\phi_*$ and $\beta$ using $\phi' _{*,{\rm in}}  = - \left( 9 \beta \epsilon / 2 \right) \phi_{*{\rm in}} \,$, according to the slow roll solution of its equation of motion (see the discussion in the paragraph above Eq. (\ref{approxsolphi})).  The initial value of $\phi_*$, and $\beta$, determine the initial value $a_{\rm in}$. This, in turns, gives us the initial value of the temperature in the Einstein frame, $T_{*\,{\rm in}} = {\rm exp} \left( a_{\rm in} \right) \, T_{\rm in} \,$. We chose values of $a_{\rm in} \lta 32 \,$, to guarantee that the initial temperature in the Einstein frame is sub-Planckian. In each case, the evolution is concluded when the temperature in the Jordan frame reaches $200 \, {\rm GeV}$ ($500 \, {\rm GeV}$) in the SM (MSSM) case, so that we do cross any mass threshold. We denote by $a_{\rm out}$ the value of $a$ at this moment, and we plot in Figure \ref{fig:sm-mssmt5} the ratio $a_{\rm out} / a_{\rm in}$ vs the initial value $a_{\rm in}$.

The ratio $a_{\rm out} / a_{\rm in}$ is nearly constant for small values of $a_{\rm in}$. The reason is the following:  small values of $a_{\rm in}$ correspond to small $\phi_*$. In this case, Eq. (\ref{finaleqp}) for $\phi_*$ is approximately linear, since one can neglect $\phi'^2$ at the denominator 
\begin{equation}
\phi_*'' + \left( 1 + \frac{3}{2} \, \epsilon(T) \right) \phi_*' + \frac{9}{2} \, \epsilon(T) \, \beta \, \phi_* \simeq 0
\;\;\;,\;\;\; \epsilon \equiv \frac{1}{3} - w
\label{approxeqphi}
\end{equation}
so that the ratio $\phi_{*\,{\rm out}} / \phi_{*\,{\rm in}}$ is independent of the initial normalization of $\phi_* \,$. Since  $a \propto \phi_*^2$, this can be rephrased in the statement that $a_{\rm out} / a_{\rm in}$ is independent of the initial value of $a_{\rm in} \,$. We can also use the approximate equation
(\ref{approxeqphi}) to understand why the decrease of $\phi_*$ observed in Figure \ref{fig:sm-mssmt5} 
is so sensitive to the value of $\beta \,$.

{}From Equation~(\ref{approxeqphi}), we see that if $\varepsilon$ vanishes, the
scalar field freezes to a constant value typically within one e-fold of expansion. 
We are thus in a over-damped
regime in which the velocity is almost constant and fixed by the source term
(see below for further justification).
Then, as long as the field is slow-rolling and the deviation of
${\mathcal I}$ from 1 is small, Eq.~(\ref{approxeqphi}) leads
to
\begin{equation}\label{e.jp0}
 \left(\frac{1}{\phi_*}-\frac{9}{2}\beta^2\varepsilon\phi_*\right)d\phi_* =
 \frac{9}{2}\beta \varepsilon d\ln T.
\end{equation}
Neglecting the variation of
$\varepsilon$ in the l.h.s. of (\ref{e.jp0}) 
 (since it is suppressed with respect to the variation 
in the r.h.s.), one can easily integrate this  equation to get
\begin{equation}\label{e.jp1}
 \ln \frac{a_{\rm out}}{a_{\rm in}} - 9\beta\hat\varepsilon a_{\rm in} 
 \left( \frac{a_{\rm out}}{a_{\rm in}}-1\right)= 9\beta\bar\varepsilon,
\end{equation}
with
\begin{equation}
 \bar\varepsilon \equiv\int_{T_{\rm init}}^{T_{\rm out}}\varepsilon(T) d\ln T
   = \sqrt{\varepsilon(T_{\rm in})\varepsilon(T_{\rm out})} \ln \frac{T_{\rm out}}{T_{\rm in}}\equiv
   \hat\varepsilon \ln \frac{T_{\rm out}}{T_{\rm in}}.
\end{equation}
Equation~(\ref{e.jp1}) can be solved in terms of the Lambert $W$ function, that is the inverse function of 
$w\rightarrow w\exp w$, as
\begin{equation}
 \frac{a_{\rm out}}{a_{\rm in}} = -\frac{1}{9\beta\hat\varepsilon \, a_{\rm in}}
 W\left[-9\beta\hat\varepsilon a_{\rm in}\hbox{e}^{-9\beta\hat\varepsilon
 \left(a_{\rm in} - \ln \frac{T_{\rm out}}{T_{\rm in}} \right)} \right],
\end{equation}
where we have chosen to fix $\varepsilon$ to its geometric
mean $\hat\varepsilon= \sqrt{\varepsilon(T_{\rm init})\varepsilon(T_{\rm out})}$ in the l.h.s. of Eq.~(\ref{e.jp0}) before integrating. This expression reproduces the computation depicted in
Fig.~\ref{fig:sm-mssmt5} with great accuracy, for $\beta \lta 50 ,\,$ and $a_{\rm in} \lta 10 \,$.

Another way to grasp some insight on the evolution of the scalar field is  to parametrize the solutions of
Eq.~(\ref{approxeqphi}) as $\phi_* = {\rm const.} \times \exp{\int^p \lambda \left( p' \right) \, d p' } \,$. The function $\lambda$ must then satisfy
\begin{equation}
\lambda' + \lambda^2 + \left( 1 + \frac{3}{2} \epsilon \right) \lambda + \frac{9}{2} \, \epsilon \, \beta \simeq 0.
\end{equation}
Let us emphasize that $\epsilon >0$ in both the SM (\ref{SM}) and the MSSM (\ref{MSSM}). The equation of state $w$ is slowly evolving and close to that of radiation (see Figure \ref{fig:w-sm-mssm}). Therefore, $\epsilon'$ and $\epsilon^2$ are both much smaller than $\epsilon$. We then find the two  solutions: $\lambda_1 \simeq - 1$ and $\lambda_2 \simeq - 9 \, \beta \, \epsilon / 2\,$. This gives us two approximate solutions of the linearized equation (\ref{approxeqphi}). To understand these solutions, consider the case $\epsilon = 0$ in which the effective potential for $\phi_*$ is absent. The two solutions then give
\begin{equation}
\phi_* = C_1 \, {\rm e}^{\int^ p \lambda_1 \, d p'} + C_2 \, {\rm e}^{\int^ p \lambda_2 \, d p'} =
C_1 \, {\rm e}^{-\left(p-p_{\rm in}\right)} + C_2 = C_1 \, \frac{R_{*{\rm in}}}{R} + C_2 \;\;\;,\;\;\; \epsilon = 0.
\end{equation}
Namely, starting from generic initial conditions (arbitrary integration constants), $\phi_*$ quickly dissipates its velocity, due to the friction term in its equation of motion, and stops at the constant value $C_2 \,$. For small $\epsilon \, \beta\,$, as we are considering here, the change in the fast decaying solution can be neglected; however, the constant mode $\propto C_2$ is  now replaced by a slowly decreasing mode. This solution is a slow roll solution of the linearized equation (\ref{approxeqphi}), in the sense that the $\phi_*''$ term can be neglected (as well as the term $\propto \epsilon \ll 1$ multiplying $\phi_*' \,$). It is clear that, starting from a generic initial $\phi_*$, after a quick initial transient regime the fast decreasing mode will become negligible. Therefore, in our numerical computations, we assume that $\phi_*$ is in the slowly rolling regime initially. Therefore, as long as Eq. (\ref{approxeqphi}) is a good approximation, $9\beta \, \epsilon/2 \ll 1 \,$, and $\epsilon$ evolves sufficiently slowly, we have the approximate solution
\begin{equation}
\phi_* \left( p \right) \simeq \phi_{*,{\rm in}} \, {\rm e}^{-\frac{9}{2} \, \beta \, \int^p d p' \, \epsilon \left( p' \right)}.
\label{approxsolphi}
\end{equation}
We see that the final value of $\phi_*$ is indeed exponentially sensitive to $\beta \,$.

\begin{figure}[h]
\centerline{
\includegraphics[width=0.4\textwidth,angle=-90]{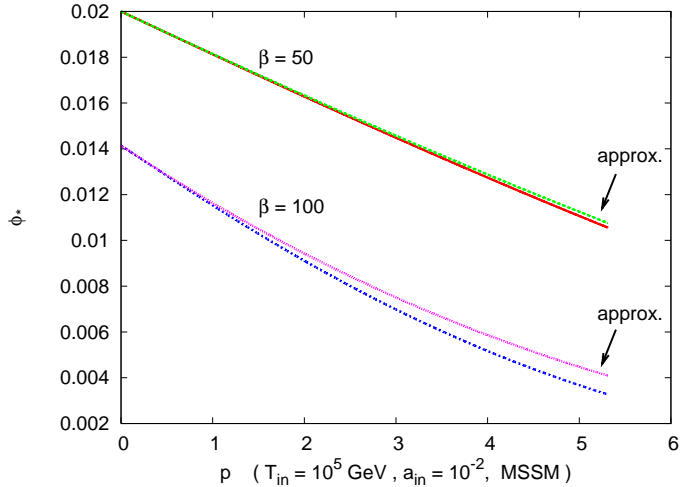}
}
\caption{A comparison of the exact and approximate solution (\ref{approxsolphi}) for $\phi_*$. The two sets of curves correspond to the same initial temperature and $a = \beta \, \phi_*^2 /2 \,$, but to two different 
values of $\beta$. 
}
\label{fig:mssm-approx}
\end{figure}

The accuracy of the approximate solution (\ref{approxsolphi}) is shown in Figure \ref{fig:mssm-approx}
for two specific choices of the parameters. We take $a_{\rm in}$ sufficiently small so that the linearized Eq. (\ref{approxeqphi}) holds, and we choose two different values of $\beta$ (we also choose $T_{\rm in} = 10^5 \, {\rm GeV}$, and we conclude the evolution when $T = 500 \, {\rm GeV} \,$, as in Figure \ref{fig:sm-mssmt5}). Accuracy of the approximate solution requires that $9/2 \, \beta \, \epsilon_{\rm in} \simeq 0.002 \, \beta \ll 1 \, $. For $\beta = 50 \,$, the two curves differ of about $2\%$ at the final moment; for  $\beta = 100$, the final discrepancy is about $20\% \,$.

\begin{figure}[t!]
\centerline{
\includegraphics[width=0.4\textwidth,angle=-90]{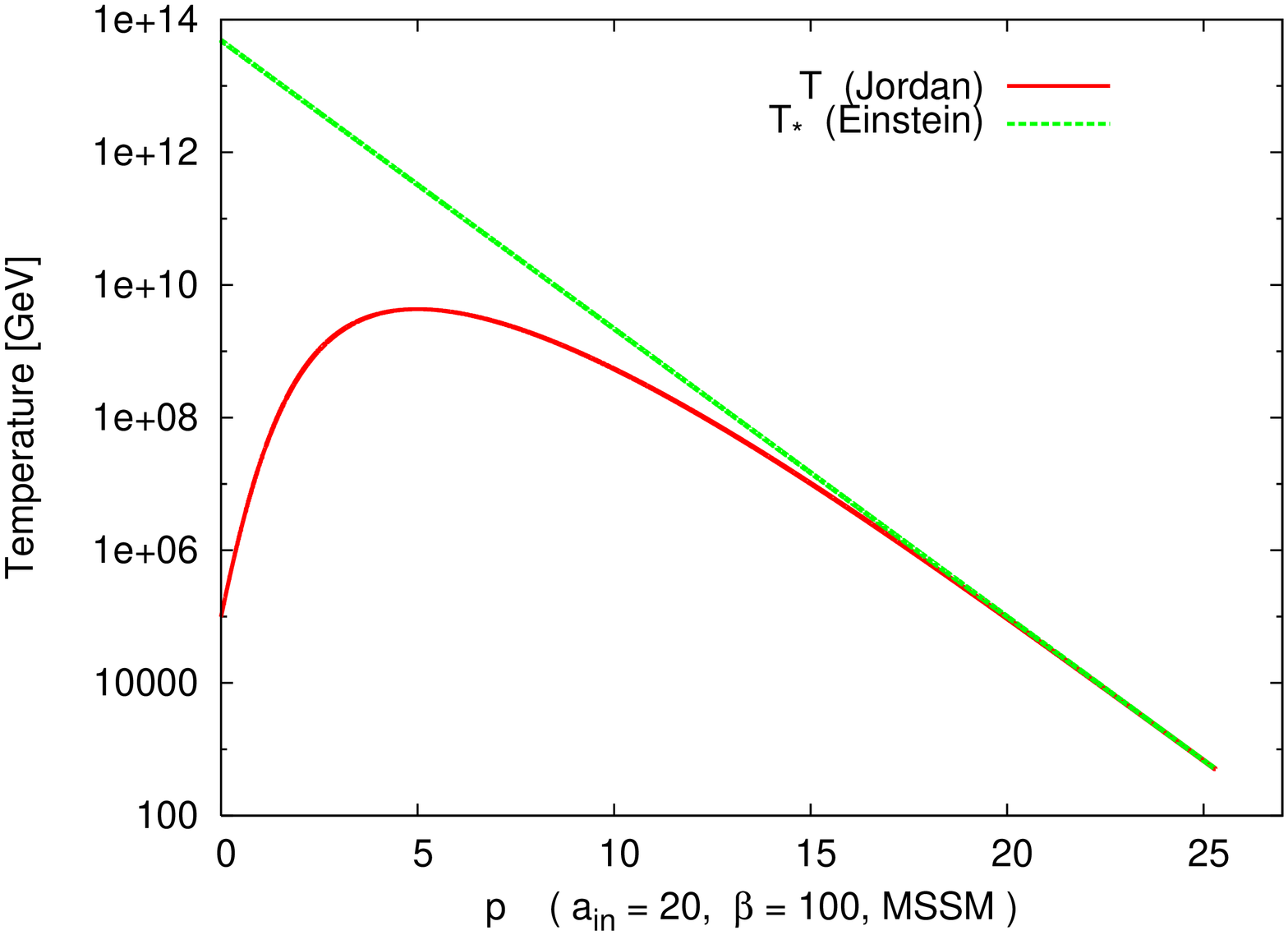}
\includegraphics[width=0.4\textwidth,angle=-90]{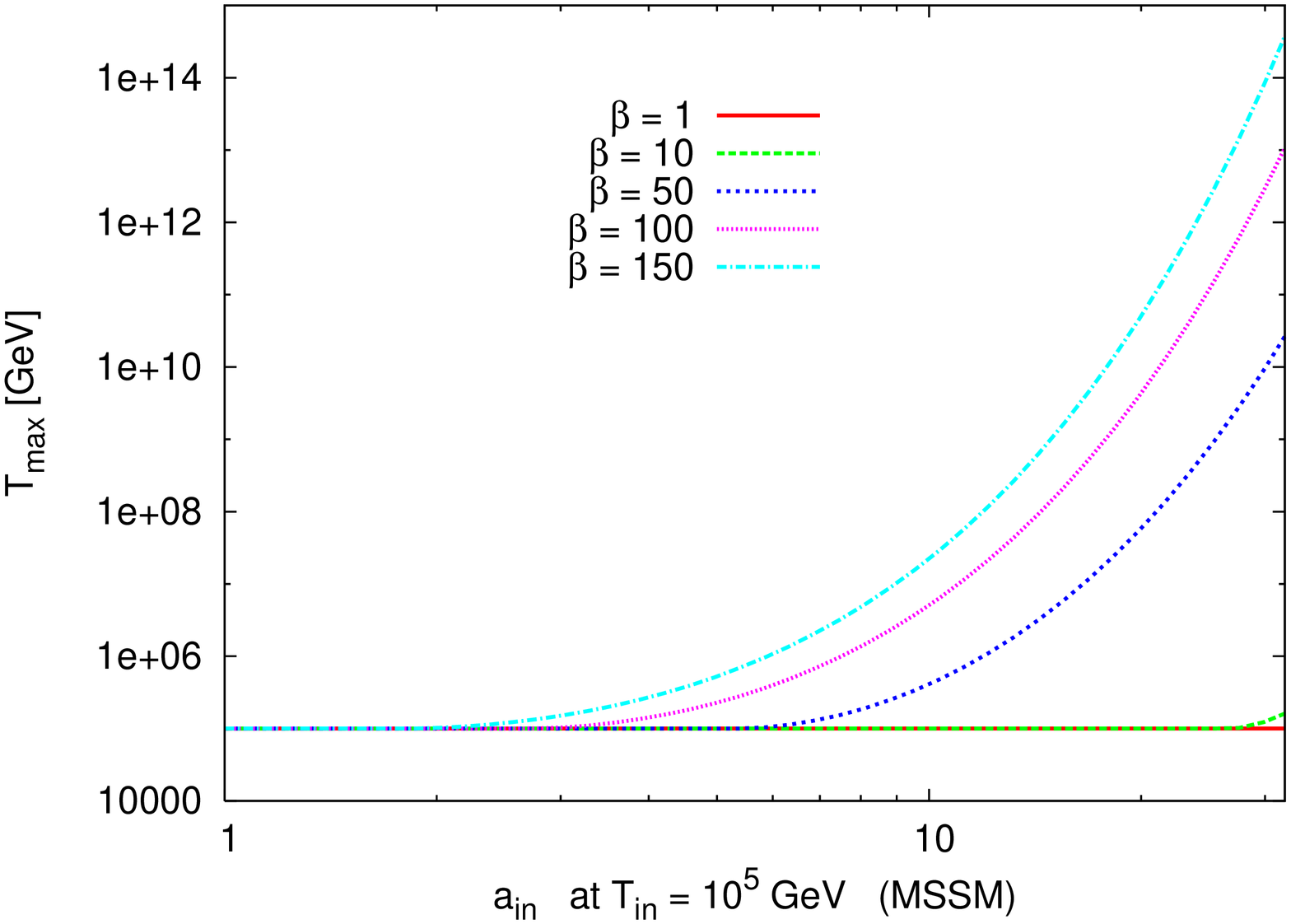}
}
\caption{Left panel: The evolution of the temperature in the Jordan and Einstein frame, for a specific choice of parameters. The temperature in the Jordan frame is increasing at the earliest times shown, it reaches a maximal value, and it then decreases. The universe contracts while $T$ increases. Notice also that the two frames coincide at late times. Right panel: The maximal temperature reached in the Jordan frame for different values of the parameters (chosen to be the same as in Figure \ref{fig:sm-mssmt5}).
}
\label{fig:mssm-temp}
\end{figure}

We also see in Figure \ref{fig:sm-mssmt5} that $a$ decreases more if we start at greater $a_{\rm in}$. Moreover, the decrease is stronger for the MSSM than for the SM. It is easier to understand this effect by discussing the evolution in the Jordan frame. As the scalar field evolves towards the origin, $a \left( \phi_* \right)$, and $A = {\rm exp} \, a$, decrease. If $A$ decreases sufficiently fast, the temperature in the Jordan frame can actually increase. To see this, recall the relation between the temperatures in the two frames, $T = T_* / A \,$, and the fact that the cosmological evolution is standard in the Einstein frame. As we discussed at the end of Section~\ref{cosmology-eqs}, the temperature in each frame is (up to ${\mathcal O} \left( 10^{-3} \right)$ corrections) inversely proportional to the scale factor in that frame. Therefore, if $A \left( \phi_* \right)$ decreases sufficiently fast, the universe contracts in the Jordan frame. As the temperature rises and then decreases again, the scalar field experiences a ``driving force'' towards the origin for a longer period of time. This causes a larger overall decrease in $\phi_*$ with respect to the cases in which $\phi_*$ (and, hence $A$) is initially small, and the evolution is standard in both frames. The suppression is more marked in the MSSM than in the SM, since the equation of state has a greater departure from $1/3$ in that case (this is particularly true at higher values of $T$, so at larger $A_{\rm in} \,$, cf. Figure \ref{fig:w-sm-mssm}).~\footnote{As we already mentioned in Section~\ref{cosmology-eqs}, the key quantity determining the departure from $w = 1/3$ is actually the ratio between the temperature and the particle mass scale, which is the same in both frames; so, this discussion can be equivalently done in the Einstein frame.} A comparison between the right panels of Figures \ref{fig:sm-mssmt5} and \ref{fig:mssm-temp} shows that indeed the ratio $a_{\rm out} / a_{\rm in}$ starts decreasing, when $a_{\rm in}$ starts out sufficiently large so that the Jordan frame temperature is initially increasing.

In the left panel of Figure \ref{fig:mssm-temp} we show the evolution of the temperature in the two frames
for some specific choice of the parameters. We choose $a_{\rm in}$ and $\beta$ sufficiently large, so that the temperature in the Jordan frame increases for some time. We verified that the temperatures are inversely proportional to their respective scale factors in each frame with a very good accuracy (more quantitatively, we found that the product $T \, R = T_* \, R_*$ increases by about $1 \%$ between the initial and final times shown). This is the reason why ${\rm log }_{10} \, T_*$ vs. $p \propto {\rm log} \, R_*$ appears as a straight line in the Figure. We know from Figure \ref{fig:sm-mssmt5} that $a$ experiences a strong decrease for this choice of parameters. Indeed, $a \simeq 0.003$ at the end of the evolution; therefore, the two frames, and the respective temperature, are (nearly) coincident at late times.

In the right panel of Figure \ref{fig:mssm-temp} we show instead the maximal temperature $T$ reached for different choices of parameters. In general, the temperature reaches a higher value for greater values of $a_{\rm in}$ and $\beta \,$. These are the cases for which $a$ shows a strong decrease, cf. the right panel of Figure \ref{fig:sm-mssmt5}. It is easy to estimate when the temperature  in the Jordan frame increases. From Eq. (\ref{finaleqT}), and recalling that ${\cal I}$ is always very close to one, we have that the temperature increases whenever $\phi_* \, {\phi_*}^\prime < - 1 / \beta \,$. Using the slow roll relation $\phi_*' \simeq - 9 \epsilon \beta/2 \,$, and our choice (\ref{our-a}) of the function $a$, this condition becomes
\begin{equation}
a \gta \frac{1}{9 \, \beta \, \epsilon} \;\;\;\Rightarrow\;\;\; T \; {\rm increases}\,.
\label{Tgrows}
\end{equation}
We can compare this estimate with the result shown in the right panel of Figure \ref{fig:mssm-temp}. Using $T = 10^5 \, {\rm GeV}$ in Eq. (\ref{MSSM}), we have an initial value of $\epsilon_{\rm in} \simeq 4.2 \times 10^{-4} \,$, and we find that the temperature is initially increasing provided that $a_{\rm in} \gta 270 / \beta \,$. More specifically, the Jordan frame temperature is initially increasing when $a_{\rm in} \gta 270,\, 27 ,\, 5.4 ,\, 2.7 ,\, 1.8 \,$, for $\beta = 1,\, 10 ,\, 50 ,\, 100 ,\, 150 \,$, respectively, and hence $T_{\rm max}$ will be greater than the initial temperature of  $T = 10^5 \, {\rm GeV}$. This is very well confirmed in the right panel of Figure \ref{fig:mssm-temp}.

It is clear that the Jordan frame temperature always reaches a maximal value, and then decreases. Indeed, even assuming that the inequality (\ref{Tgrows}) is initially satisfied, $a \propto \phi_*^2$ decreases, while the right hand side slowly increases (since $\epsilon$ slowly decreases as $T$ grows). $T_{\rm max}$ is reached whenever the inequality stops holding. Therefore, $a_{\vert T_{\rm max}} \simeq 1 / \left( 9 \, \beta \, \epsilon \right)$, and $\phi_{*\vert T_{\rm max}} \simeq \sqrt{2/\epsilon} / \left( 3 \, \beta \right) $ when the Jordan frame temperature reaches its maximal value. If we also have an analytical expression for $p$ at this moment, we can then use Eq. (\ref{solT}) to obtain an analytical expression for $T_{\rm max} \,$. The redshift $p$ can be related to the value of $\phi_*$ that we have just obtained through the approximate solution (\ref{approxsolphi}). Unfortunately, $\epsilon$ varies with temperature, so that we cannot straightforwardly invert Eq.~(\ref{approxsolphi}) to find $p$ as a function of $\phi_*$. To obtain an estimate, we simply assume that $\epsilon$ is constant, so that
\begin{equation}
{\rm e}^{-p} \sim \left( \frac{\phi_*}{\phi_{*,{\rm in}}} \right)^{\frac{2}{9 \beta \epsilon}} \;\;\;,\;\;\; \epsilon \equiv {\rm const.}
\end{equation}
By inserting our estimate for $\phi_{*\vert T_{\rm max}} $, and the relation between $\phi_{*,{\rm in}}$ and $a_{\rm in}$ in this expression, we obtain an analytic expression for $p$ at  $T_{\rm max}$. Inserting all of this into Eq.~(\ref{solT}), we find
\begin{equation}
T_{\rm max} \sim T_{\rm in} \, {\rm e}^{a_{\rm in}} \, \left( \frac{1}{9 \, e \, \beta \, a_{\rm in}} \right)^{\frac{1}{9 \, \beta \, \epsilon}} \;\;\;,\;\;\; \epsilon \equiv {\rm const.}
\label{estimateTmax}
\end{equation}
This estimate assumes that the inequality (\ref{Tgrows}) is initially satisfied, so that $T$ is growing initially (if this is not the case, $T_{\rm max}$ is simply $T_{\rm in}$). A comparison with the exact values for $T_{\rm max}$ shown in the right panel of Figure \ref{fig:mssm-temp} indicates that the estimate (\ref{estimateTmax}) is accurate provided $T_{\rm max}$ is about $2-4$ orders of magnitude greater than $T_{\rm in}$ (the precise value depends on the choice of $\beta$). For a larger span of temperatures, $\epsilon$ can no longer be assumed to be constant, and the estimate (\ref{estimateTmax})
loses its accuracy.

We conclude this Section with a few comments on the effect of starting from an arbitrary value of $\phi_*'$ (rather than starting from the slow roll value obtained from Eq. (\ref{approxsolphi})). One way to ensure that $\phi_*$ is always in the slow roll regime is to imagine that the field responsible for inflation is coupled differently to $\phi_*$ than to matter fields. For instance, if the inflaton is decoupled from $\phi_*$ in the Einstein frame, $\phi_*$ has no driving force during inflation. Due to friction, $\phi_*$ reaches a constant value, and inflation proceeds as in GR.  The effective potential for $\phi_*$ is generated at reheating, when the inflaton decays to the matter fields which are coupled to the scalar. In this scenario, it is reasonable to assume that $\phi_*$ will settle in the slow roll regime as reheating proceeds and the effective potential is being created. However, without a complete theory, it is legitimate to assume any value for $\phi_*'$ at the start of our simulations. 

Wen considering arbitrary initial velocity, we have however to restrict the choice to $\vert \phi_*' \vert < \sqrt{3}$. Indeed, the first equation of (\ref{eq-einstein}) can be rewritten as: $H_*^2 ( 3 - \phi_*'^2 ) = 8  \pi G_* \rho_*$. Solutions with $\vert \phi_*' \vert > \sqrt{3}$ require $\rho_* < 0$ and are therefore unphysical. Moreover, the two branches of solutions with $\vert \phi_*'\vert$ greater or smaller than $\sqrt{3}$ are mathematically disconnected. For the physically meaningful case, the effect of 
the $3 - \phi_*'^2$ prefactor in Eq. (\ref{finaleqp}) can be neglected in the present qualitative discussion, so that with our choice of $\alpha$, we simply have the equation of a scalar in a quadratic potential subject to friction. The friction will damp the initial velocity, so that, after an initial transient regime, the field proceeds as in the cases we studied. We can however imagine the situation in which in this initial phase the field is actually moving away from the origin (it will then stop, due to the friction term, and move back towards the minimum at $\phi_* = 0 \,$). In general, as is it clear from Eq. (\ref{solT}), the temperature $T$ in the Jordan frame decreases faster as $\vert \phi_* \vert $ increases, while it decreases more slowly (or even increases) when $\vert \phi_* \vert$ decreases.

\section{Discussion}

\label{discussion}

In this work, we have studied the dynamics of CCSFs in a radiation dominated universe. We have shown that there are cases in which the evolution of the scalar  field can have a very strong impact on cosmology. Despite the conformal coupling, the scalar field has a source term given by the conformal anomaly, which is ultimately related to the fact that particles in the thermal bath are interacting (this leads to running coupling constants, and to $T^\mu_\mu = \rho - 3 \, P \neq 0$ in the thermal bath).
We computed the leading value for the source term both in the SM and the MSSM.  We then focused on the effect that it can have in the context of  scalar tensor theories of gravity, in which the scalar field $\phi_*$ is conformally coupled to matter (by ``matter'', we mean any SM or MSSM field). We specifically considered scalar tensor theories which have GR as an  attractor.  In such theories, the additional coupling between matter mediated by $\phi_*$ has a strength which is function of the scalar field itself, and which vanishes for some given value of $\phi_*$. One simple way to formulate the theory is to impose a standard ($\phi_*$-independent) action for the matter fields, but a nonstandard curvature term, with $M_p^2 \propto A^{-2} \left( \phi \right)$ - this is known as the Jordan frame. Then, the strength of the new interaction is controlled by $\alpha \equiv d \, {\rm ln } \, A / d \phi_*$. If present, matter fields with $T^\mu_\mu \neq 0$ tend to drive $\phi_*$ to a value for which $\alpha$ vanishes, and GR is recovered \cite{dnord}. The efficiency of this mechanism increases with increasing $T^\mu_\mu$, and with increasing curvature of ${\rm ln } \, A$ close to the minimum. The latter quantity is typically denoted by $\beta \,$; our explicit study is performed for the simplest case in which $\beta$ is constant ($\phi_*$-independent). However, we expect that the qualitative features that we find are general. As we remarked, precision tests of GR impose that $\alpha_0^2 \lta 10^{-5} \,$ and $\vert \beta_0 \, \alpha_0^2 \vert \lta {\mathcal O} \left( 10^{-3} \right)$ today. This does not implies that $\beta$ should be small (indeed, increasing $\beta$, leads to a smaller value for $\alpha$ today). Moreover, this leaves significant room for departures from GR at earlier cosmological eras.

As we showed in Figure \ref{fig:w-sm-mssm}, the trace anomaly gives a small departure from the noninteracting result, $1/3 - w = {\mathcal O } \left( 10^{-4} - 10^{-3} \right)$, at temperatures much greater than the mass of any SM or MSSM particle. Nonetheless, in some cases this can have a strong impact on the evolution of $\phi_*$, and on the whole cosmology. The most obvious effect is that $\alpha$ decreases during this era, due to the mechanism of attraction towards GR~\cite{dnord}. The amount of the decrease is depicted in Figure \ref{fig:sm-mssmt5}, where we see that the effect is strongly sensitive to $\beta$ and to the initial value of $A$. Previous cosmological studies of such theories (see for instance Ref.~\cite{dp,couv}) only considered the   attraction towards GR during (i) the matter dominated era ($w=0$), or (ii) in the radiation dominated era, whenever the temperature crosses any mass threshold (this results in $w \neq 1/3$ when the temperature is close to the mass of any particle in the thermal bath). Our work refers to the previous cosmological epoch, and therefore, our final values should be understood as the initial values for such studies. Since $\alpha$ in general decreases for $T^\mu_\mu \neq 0$, we generally extend the region of allowed initial conditions for such models. The amount of the extension depends on the initial temperature, as well as on the initial values of $\beta$ and $\alpha \,$. To provide some ``benchmark'', based on the example we studied, we found that our effect is non-negligible whenever $\beta \gta 10$, and very strong for $\beta \sim 50 - 100$ or greater.

A second, less obvious, effect of the conformal anomaly is that it can lead to a contracting phase of the universe, followed by a regular bounce, and standard expansion. 
Similar (bouncing) behaviour has been postulated as a possible alternative to inflation and in other cosmological contexts as well \cite{mg1}. 
The cosmological evolution appears as standard in the Einstein frame (the frame in which the gravity term is regular, but the metric felt by matter contains the function $A$), with $\phi_*$ contributing to the energy density and pressure as a standard scalar field. Therefore, the scale factor of the universe $R_*$ in the Einstein frame always increases. The temperature $T$ in the Jordan frame is related to the temperature $T_*$ of the Einstein frame by $T = T_* / A \left( \phi_* \right)$. Moreover, from the
conformal transformation we have  $R = R_* \, A \left( \phi_* \right) \,$. If the evolution of $\phi_*$ leads to a sufficiently fast decrease of $A$, then $R$ can decrease, even if $R_*$ is increasing. This leads to a temporary phase in which the universe contracts, and $T$ increases. Also in this case, the effect is more marked for large $\beta$ and $A_{\rm in} = e^{a_{\rm in}}$. This can be seen in Figure \ref{fig:mssm-temp}, where we show the maximal value reached by $T$ when the universe stops contracting and bounces, for different values of $\beta$ and $A_{\rm in}$, but for the same initial temperature $T= 10^5 \, {\rm GeV}$. We see that $T$ can increase by several orders of magnitude before the bounce.

Before discussing the possible consequence of the bounce, it is useful to make a few clarifications. The first is that whether the universe is contracting or expanding is obviously a frame independent question, whose resolution is most immediately seen from the behavior of $R$ rather than $R_*$. To see whether the universe expands or contracts, cosmological distances need to be compared with a ``ruler''. It is convenient to choose the Compton wavelength of any particle, $\lambda \propto m^{-1}$ (where $m$ is the mass) as the ruler~\cite{Campbell:1990de}. In the Jordan frame, $\lambda$ is constant (since the action of matter is independent of $\phi_*$). Therefore, the universe expands/contracts whenever $R$ increases/decreases. It is easy to show that the mass of a particle $m_*$ in the Einstein frame is related to the mass $m$ in the Jordan frame by $m_* = m \, A \left( \phi_* \right) \,$. Therefore, the ratio between the scale factor and the Compton wavelength is the same in both frames. The second clarification is that, in principle, one can have a regular bounce in a scalar tensor theory of gravity even if one disregards the trace anomaly. Since  $T = T_* / A \left( \phi_* \right)$, one ``simply'' needs to arrange for a sufficiently fast decrease of $A$ for some period of time. For instance, it may be possible to obtain this with a sufficiently high initial velocity for $\phi_* \,$. Here, we have neglected any transient phase due to the initial value of $\phi_*' \,$. From the evolution equation for $\phi_*$ one can easily see that the initial velocity is dissipated in about one e-fold of expansion (in the Einstein frame). After that, the evolution of $\phi_*$ is simply dictated by its coupling to matter. In the early universe, neglecting the conformal anomaly leads to $\phi_*' = 0 \,$. Including (as one must) the effect of the conformal anomaly leads to the bounce studied here.

Regarding the possible consequences of the bounce, it is tempting to wonder whether it could lead to a solution of some of the standard cosmological problems, such as the horizon problem. If the universe could contract and remain small for a long enough time, one could hope that the propagation of light in that time could result in a much larger horizon than achieved in a  standard cosmology. We have verified that this is not the case. This can be most easily understood in the Einstein frame, in which the
cosmological evolution is standard, leading to a horizon size which is nearly coincident with the standard result. We can rephrase this by stating that, in the Jordan frame, the universe remains small for a negligible amount of time.

The fact that the universe bounces and reaches a maximal temperature in the Jordan frame can have instead very interesting consequences for the thermal production of particles and topological defects. Particle physics processes are most immediately computed in the Jordan frame, in which the particles masses and interactions are independent of $\phi_* \,$. Therefore, computations of thermal scatterings and symmetry restorations are standard in the Jordan frame. We can arrange $T_{\rm max}$ to remain sufficiently small, so that the thermal production of unwanted relics is sufficiently suppressed, and no symmetries which would result in unwanted defects are restored. This could happen even if the temperature $T_*$ in the Einstein frame is high, see for instance the left panel of Figure \ref{fig:mssm-temp}. Provided $A_{\rm in}$ is sufficiently high, so to provide an initial large hierarchy between the Einstein and Jordan frame temperatures, and $\beta$ is not too large, so that $T$ does not increase too much during the evolution (see the right panel of Figure \ref{fig:mssm-temp}), one could even start with the highest possible value $T_* \approx G_*^{-1/2}$ in field theory (i.e., before entering in the quantum gravity regime), without $T$ ever conflicting with any cosmological bound, such as that arising from the gravitino problem. The presence of a bounce could also have consequences for wanted relics, namely baryons and dark matter, if $T_{\rm max}$ is close to the decoupling temperature of dark matter or to the mass of a particle responsible for the baryon asymmetry (for instance, the lightest right handed neutrino in leptogenesis).~\footnote{Limits on scalar tensor theories from the dark matter relic abundance were obtained for instance in 
Ref.~\cite{Catena:2004ba} by considering only the evolution of $\phi_*$ due to the mass thresholds.}

We stress that, while the present study has been focused on scalar tensor theories of gravity, the effect we have studied is relevant for any CCSF. Another interesting application is the evolution of the dilaton $\Phi$ in string models. The dilaton is also conformally coupled to matter, and it posses a mechanism analogous to the attraction to GR we have studied, namely the least coupling principle 
\cite{dpol}. Also in that context, one can go to the Einstein frame, in which the gravity term is standard, and canonically normalize the matter fields. Neglecting the interactions between different fields, the dilaton then enters in the expression for the masses of the various fields. This results in the following equation of motion for the dilaton \cite{dpol}
\begin{eqnarray}
\Phi'' &=& - \frac{3 - \Phi'^2}{2} \left[ \left( 1 - w \right) \Phi' + \sum_A \alpha_A \frac{\rho_A - 3 \, P_A}{\rho} \right], \nonumber\\
\alpha_A &\equiv& \frac{d \, {\rm ln } \, m_A \left( \Phi \right)}{d \Phi}
\end{eqnarray}
where prime again denotes derivative with respect to $p$, the index $A$ denotes any matter field, and $\rho$ is the total matter energy density, $\rho = \sum_A \rho_A$. This equation is analogous to eq. (\ref{finaleqp}) for $\phi_*$. If there was only one matter field, the expression would actually be identical. Indeed, since in scalar tensor theories of gravity $m_*= A \left( \phi_* \right) \, m$, with a $\phi_*-$independent Jordan frame mass $m$, the coupling $\alpha$ entering in Eq. (\ref{finaleqp}) and defined in Eq. (\ref{defalpha}) can also be expressed as $\alpha = d \, {\rm ln } \, m_* / d \phi_* \,$. For the dilaton case, there is the additional complication that the functional dependence $m_A \left( \Phi \right)$ can be different for different matter fields. However, if all the functions $\alpha_A$ vanish for one specific value of $\Phi$, then the matter fields drive the dilaton to that value, in the same way as $\phi_*$ is driven to a zero of $\alpha$~\cite{dpol}. For this value, the dilaton decouples from matter. Due to the trace anomaly we have discussed, the dilaton will be attracted to that value also during the radiation dominated era. This can possibly result in larger range of initially allowed values for $\Phi$ than what one would have neglecting the trace anomaly (as for the case we studied, the size of this effect depends on the specific functional forms of $m_A \left( \Phi \right) \,$).

As we mentioned, the consequences of the conformal anomaly are more sizeable for large $\beta$ and $A_{\rm in}$. In most of the examples shown, we started from $T_{\rm in} = 10^5 \, {\rm GeV}$, and we considered $A_{\rm in}$ up to $\sim 10^{14} \,$, so that, when this limit is saturated, $T_{*,{\rm in}}$ is Planckian. It is tempting to relate the hierarchy between the electroweak and the Planckian scale to 
a large conformal factor. This is precisely what is done for the dilaton in the context of the so called little string theory at the TeV \cite{little}. Consider the string theory expression for the Planck mass $M_p \sim M_s^8 \, V_6 / g_s^2 \,$, where $g_s$ is the string coupling, $M_s$ the string scale, and $V_6$ the volume of the six-dimensional internal space. Assuming that the fundamental scale is TeV, a large $4$d planck mass can be obtained by a large volume $V_6$ \cite{add}. Alternatively, one can assume that both $M_s$ and $V_6$ are set by the TeV scale, and that the large Planck mass is due to a very small string coupling $g_s \sim 10^{-16} \,$. From the expansion series $g_s = {\rm e}^\Phi + \dots$, this then requires a large negative value for the dilaton. Therefore, it is argued in the last of Ref.~\cite{little} that the
large value of the Planck mass today can be related to the runaway of the dilaton, assuming that nonperturbative terms in the potential stabilize $\Phi$ at the desired value (this work also discusses limits on the theories imposed by deviations from GR). In our context, we can express the Planck mass 
(i.e., the coefficient of $R$ in the action) in the Jordan/string frame as (cf. Eq.~(\ref{act-jordan}))
\begin{equation}
M_p^2 = \frac{1}{G_*} \, \frac{1}{A^2 \left( \phi_* \right)} =  \frac{1}{G_* \, A_{\rm in}^2} \, \frac{A_{\rm in}^2}{A^2 \left( \phi_* \right)} \equiv {\rm TeV}^2 \, {\rm e}^{\beta \, \left( \phi_{*,{\rm in}}^2 - \phi_*^2 \right)}
\label{mpain}
\end{equation}
Namely, we assume that the Planck mass is initially at the TeV scale, and it is then driven to a large value by the evolution of $\phi_*$. Therefore, the last term in (\ref{mpain}) plays the same role as $1/g_s \left( \Phi \right)$ in the string theory expression. Although our choice of $A \left( \phi_* \right)$ was dictated by simplicity, one interesting extension of our study would be to consider other functional forms which reproduce the string dilaton case.

To conclude, modifications of GR can result in interesting nonstandard cosmologies at early times, when 
such modifications are less constrained. In this work, we considered the effect of the trace anomaly at temperatures higher than the electroweak scale. Somewhat surprisingly, we saw that that the universe can experience a regular bounce in this context. As we approach this early time regime, it is natural to study how this effect can be reconciled within a complete cosmological framework, where the standard cosmological problems are solved, and the primordial perturbations are obtained in agreement with observation. 
We believe that this issue deserves further investigation.

\vspace{.2 cm}

\section*{Acknowledgements}
We are thankful to N.~Kaloper and J.~I.~Kapusta for very useful discussions.
The work of JC, KAO, and MP was supported in part by DOE grant
DE-FG02-94ER-40823 at the University of Minnesota.

\end{document}